\documentclass[aps,prx,twocolumn,reprint,superscriptaddress,floatfix,footinbib]{revtex4-2}
\usepackage{comment}
\usepackage{graphicx}
\usepackage{amsmath}
\usepackage{setspace}
\usepackage{braket}
\usepackage{mathtools}

\usepackage{natbib,mathrsfs,accents}
\usepackage[dvipsnames]{xcolor}
\definecolor{myblue}{named}{MidnightBlue}
\definecolor{mygreen}{RGB}{0,192,0}
\usepackage[colorlinks=true,citecolor=myblue,linkcolor=myblue,urlcolor=myblue]{hyperref}
\usepackage{siunitx}
				
\usepackage[caption=false]{subfig}

\makeatletter
\def\thickhline{%
  \noalign{\ifnum0=`}\fi\hrule \@height \thickarrayrulewidth \futurelet
   \reserved@a\@xthickhline}
\def\@xthickhline{\ifx\reserved@a\thickhline
               \vskip\doublerulesep
               \vskip-\thickarrayrulewidth
             \fi
      \ifnum0=`{\fi}}
\makeatother

\usepackage{url}

\begin{document}

\title{Time-delayed single satellite quantum repeater node for global quantum communications}

\author{Mustafa Gündoğan}
\email{mustafa.guendogan@physik.hu-berlin.de}
\affiliation{Institut f\"{u}r Physik and IRIS Adlershof Humboldt-Universit\"{a}t zu Berlin, Newtonstr. 15, Berlin 12489, Germany}
\author{Jasminder S. Sidhu}
\affiliation{SUPA Department of Physics, University of Strathclyde, Glasgow, G4 0NG, UK}
\author{Markus Krutzik}
\affiliation{Institut f\"{u}r Physik and IRIS Adlershof Humboldt-Universit\"{a}t zu Berlin, Newtonstr. 15, Berlin 12489, Germany}
\affiliation{Ferdinand-Braun-Institut (FBH), Gustav-Kirchoff-Str.4, 12489 Berlin}
\author{Daniel K. L. Oi}
\email{daniel.oi@strath.ac.uk}
\affiliation{SUPA Department of Physics, University of Strathclyde, Glasgow, G4 0NG, UK}
\affiliation{Walton Institute for Information and Communication Systems Science, South East Technological University, Waterford, X91 P20H, Ireland}

\begin{abstract}
Global-scale quantum networking faces significant technical and scientific obstacles. Quantum repeaters (QRs) have been proposed to overcome the inherent direct transmission range limit through optical fibre. However, QRs are typically limited to a total distance of a few thousand kilometres and/or require extensive hardware overhead. Recent proposals suggest that strings of space-borne QRs with on-board quantum memories (QMs) are able to provide global coverage. Here, we propose an alternative to such repeater constellations using a single satellite with two QMs that effectively acts as a time-delayed version of a single QR node. Using QKD as a benchmark, we estimate the amount of finite secure key generated and demonstrate an improvement of at least three orders of magnitude over prior single-satellite methods that rely on a single QM, while simultaneously reducing the necessary memory capacity similarly. We propose an experimental platform to realise this scheme based on rare-Earth ion doped crystals with appropriate performance parameters.
\end{abstract}

\maketitle

\section{Introduction}
\label{S:1}

\noindent

Long-distance ($>10^3$~km) quantum entanglement distribution will be crucial for the development of global networked quantum computers, sensors, positioning, navigation and timing, as well as for fundamental tests of physics. The main scientific and technical challenge is the high loss suffered by directly transmitted quantum signals that constrains the range and rate of entanglement distribution. Unlike in classical communications, deterministic amplification of quantum states is prohibited by quantum mechanics~\cite{Dieks1982, Wootters1982}. Currently, fibre-based long-distance quantum communication experiments are limited to around a few hundred to a thousand kilometres~\cite{Chen2020}, made possible by new techniques such as twin-field (TF) quantum key distribution (QKD)~\cite{Lucamarini2018,liu2023experimental} and developments in low-loss fibre and low-noise single photon detectors. However, going beyond $\sim10^3$~km requires alternate approaches, such as quantum repeaters (QRs) or free-space channels via space-based platforms.

Actively corrected QRs have theoretically shown to be capable of reaching global distances albeit with enormous technical overhead: repeater nodes, each of which contains a small-scale quantum processor, must be separated by very short distances, i.e 1-2~km~\cite{Muralidharan2014}, due to sensitivity to link loss. QRs based on heralded generation of entanglement~\cite{Childress2006,Sangouard2011} can have nodes separated by several tens of kilometres but their total range is limited to around a few thousand kilometres~\cite{Duan2001, Simon2007, Sangouard2011, Vinay2017}.

The use of free-space channels can also extend the direct-transmission limit, fibre exponential loss scaling is replaced with the (mainly) inverse square loss scaling of free-space propagation. Recently, the Micius satellite~\cite{Lu2022} demonstrated milestones such as ground-space teleportation~\cite{Ren2017}, QKD with entangled photons across 1120~km~\cite{Yin2020}, intercontinental QKD operated in trusted node~\cite{Liao2018} and the integration of satellite links into long-distance, trusted node ground networks~\cite{Chen2021}. These groundbreaking achievements are however limited by line-of-sight, the connection distance $d$ between two ground stations is limited by the requirement to be in simultaneous view of the satellite ($d\le2000$~km for altitude $h=500$~km) unless the satellite operates as a trusted node~\cite{Liao2018,Vergoossen2020}. 

Fully global ($d>10^4~\text{km}$) coverage with satellites has been proposed by several groups. An initial proposal was a hybrid, space-ground QR system~\cite{Boone2014} where quantum memories (QMs) were located in ground stations. This scheme was recently extended towards fully satellite-based QRs~\cite{Liorni2021, Gundogan2021} where the QMs are located onboard satellites~\cite{DaRos2023}, eliminating intermediate trans-atmospheric quantum links. These works demonstrated that entanglement distribution across the whole globe would be possible with a network of satellites equipped with QMs and entangled photon pair sources. It was shown that a storage time of around $<$1~s would be sufficient to reach global distances~\cite{Liorni2021, Gundogan2021} whereas intercontinental distances of $>8000$~km would be possible with memory times of around 100~ms~\cite{Gundogan2021, Wallnofer2022}.

An alternative to multiple QR nodes is to physically transport~\cite{Transport2020, Zhou2022} entangled qubits, given sufficiently long qubit coherence times. This could be achieved via active quantum error correction~\cite{Devitt2016} or by with ultra-long lifetime (ULL) QMs~\cite{Wittig2017, Bland-Hawthorn2021}. Here, we propose a time-delayed version of a single-node quantum repeater~\cite{Luong2016, Trenyi2020, Langenfeld2021} that can be implemented with a single orbiting satellite carrying an ULL QM (QM1) in combination with a shorter lived ($\sim 10s$~ms) QM (QM2). The addition of the lower requirement QM2 provides a feasible route towards dramatic improvements in secure key generation over QM1 alone. Using QKD as a benchmark for quantum communication~\cite{Tysowski2018_QST}, our scheme extends the performance and reduces the hardware requirements of a previous related proposal~\cite{Wittig2017} by several orders of magnitude when taking into account finite-block size effects.

This manuscript is organised as follows: we first outline the protocol with two quantum memories, highlighting differences from a previous single-memory scheme; we present finite key analysis of this scheme and derive required memory performance; finally, we discuss possible implementations of this scheme and provide an experimental guideline.

\section{2-QM Time-Delayed QR Protocol}

\begin{figure*}
    \begin{center}
 \includegraphics[width = \textwidth]{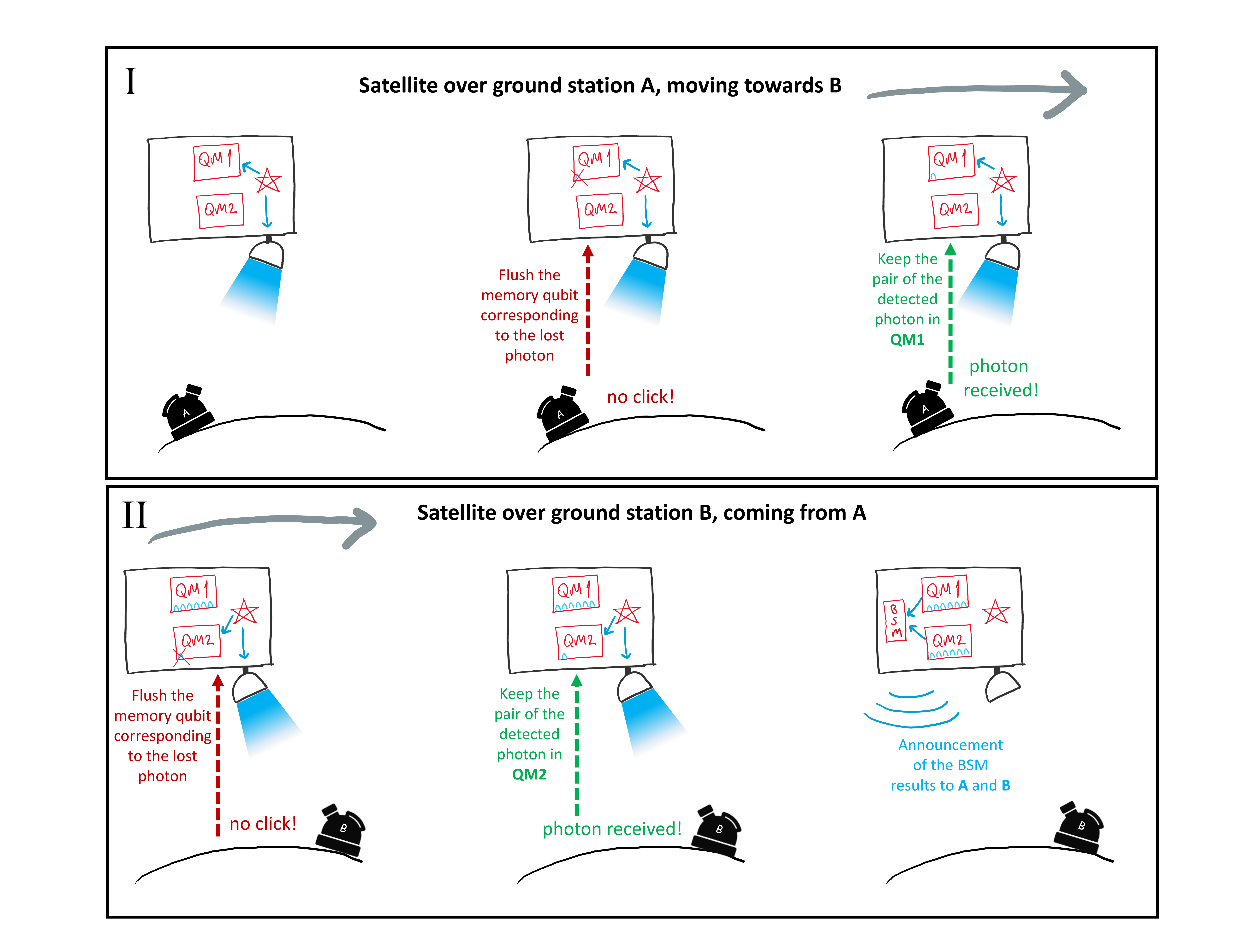}
    \caption{2-QM Time-Delayed Protocol. (I) One photon of the entangled pair is stored in the long-lived QM1 whereas its partner is sent to ground station A which informs the satellite if it was received. If ground station A does not receive the photon, the respective QM1 mode is reset, and entanglement distribution is reattempted, otherwise the stored photon is kept in QM1. This continues until all storage modes of QM1 are full, or the transmission window is exceeded. (II) The satellite continues in its orbit and passes close to ground station B. We utilise QM2 in a similar process as in (I) to establish shared entanglement with ground station B. Entanglement swapping is now performed by applying a BSM to the stored photons in QM1 and QM2 whose result is broadcast to ground stations A and B who may perform local unitary corrections.}
    \label{fig:2memory}
    \end{center}
\end{figure*}

\noindent
Global quantum communications using a low-Earth orbit (LEO) satellite equipped with an ULL QM and an entangled photon pair source has been previously proposed~\cite{Wittig2017}. The source first sends one of the photons in each entangled pair to ground station A and the other half is stored in the on-board QM. The satellite then continues in its orbit and stored photons in the QM are retrieved and sent to ground station B as it flies over. Our scheme instead supplements the ULL QM with a second shorter-lived QM. QM1 needs to have $\tau_{QM1}>1$~hour with a high multimode capacity whereas QM2 only needs $\tau_{QM2}\sim2L/c$, where $L$ is the range between the satellite and the ground station. Using a single QM limits the keylength scaling to $\eta_{ch}^2$ whereas a second QM enhances the scaling to $\sim\eta_{ch}$, where $\eta_{ch}$ is the average single channel loss, all else being equal. Our scheme can be regarded as the time delayed version of a single quantum repeater node~\cite{Luong2016, Trenyi2020, Langenfeld2021} that enhances the achievable key rates and tolerable losses and eliminates the requirement for the two ground stations to be in simultaneous view.

We illustrate our scheme in Fig.~\ref{fig:2memory}. The protocol begins with the start of the satellite pass over ground station A and operating its entangled photon-pair source (EPS, rate $s$). One photon in each pair is stored in QM1, the other photon is sent to ground station A through the space-ground channel. If this transmitted photon is lost, then the corresponding stored photon in the QM1 is erased, else the stored photon is kept if ground station A indicates successful reception. For QKD, the received photon is measured (as in BBM92~\cite{bennett1992quantum}) or, more generally, the ground stations could store the received photons in a QM if entanglement was required instead.

After the first overpass, the satellite continues in its orbit and the source again starts emitting entangled photon pairs when passing over ground station B. One photon of each pair is sent to ground station B whereas the other photon is stored in QM2. If ground station B successfully receives the transmitted photon, then the corresponding photon from QM2 is immediately retrieved together with a photon stored in QM1 and entanglement swapping performed by a Bell state measurement (BSM). The result of the BSM is then broadcast for local unitary corrective operations~\cite{Luong2016, Trenyi2020}. Although we consider a QM paired with an entangled photon pair source~\cite{Simon2007, Rakonjac2021}, the same protocol can be realised with a DLCZ-type memory, where the QM can emit a single photon entangled with its internal atomic states~\cite{Duan2001}. 

\section{Key rate analysis}

\begin{figure}[t]
\centering
  \begin{tabular}{@{}cc@{}}
    \includegraphics[width=0.95\columnwidth]{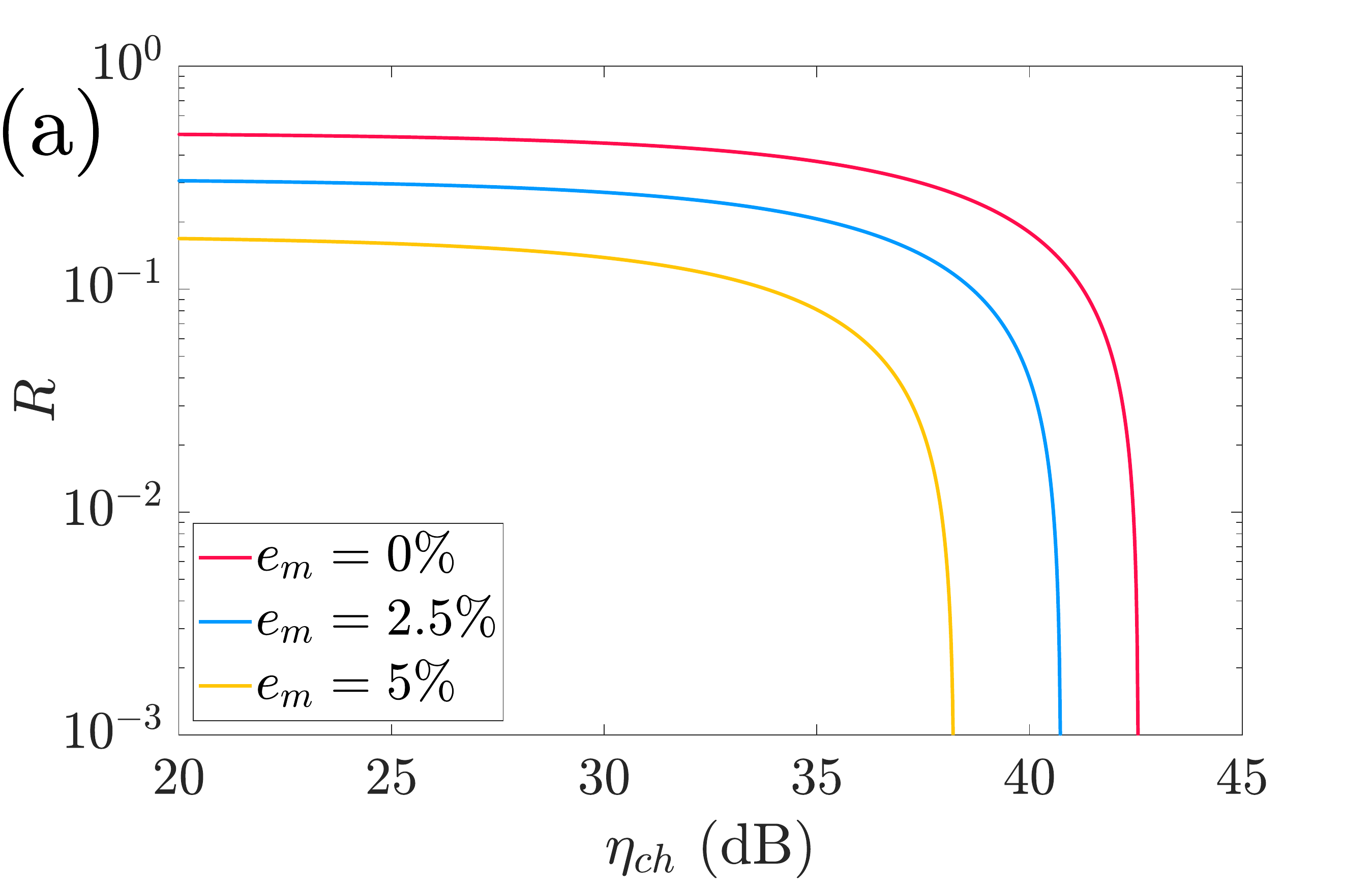}\\
    \includegraphics[width=0.95\columnwidth]{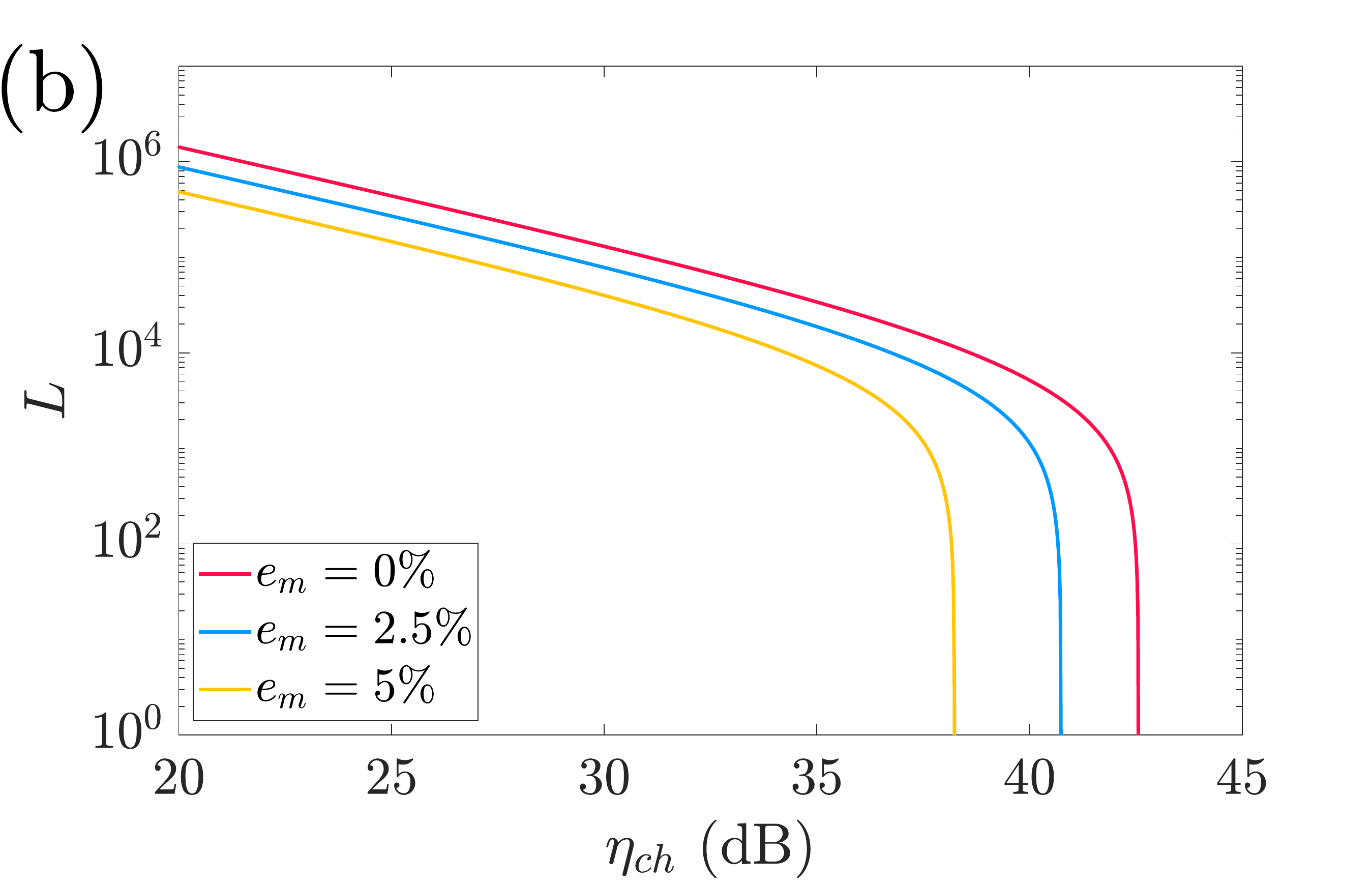}\\
  \end{tabular}
  \caption{2-QM QKD Performance. Overpass time $240$~s consistent with a typical LEO overpass, $5$~MHz source rate, average single downlink channel loss $\eta_{ch}$. a) Secure key rate per received pair,  \textit{R}; b) Secure key length, \textit{L}. Model parameters are given in Table~\ref{tab:parameters}.} \label{Fig2} 
\end{figure}

Quantum key distribution rates are often calculated in the asymptotic limit~\cite{Pirandola2020} assuming that statistical uncertainties can be neglected due to sufficiently large blocks sizes. However, this approximation may be invalid for limited duration overpasses, hence finite block size effects should be considered~\cite{sidhu2021advances,islam2022finite,Sidhu2022,sidhu2023finite,sidhu2023satellite}. Here, for comparative purposes, we follow the approach of Ref.~\cite{Yin2020} for the analysis of BBM92 with symmetric basis choice due to its relative simplicity. Although this analysis was refined later in Ref.~\cite{Lim2021}, we assign a tighter security parameter to maintain the security of finite keys by appropriately accounting for the cost of parameter estimation.

Considering the successfully entanglement-swapped pairs shared by ground stations A and B, the finite key length in the $Z$ basis is then given by,
\begin{align}
\begin{split}
    L_{Z} =& n_{Z}-n_{Z} h\left[\frac{e_{X}+\sqrt{\frac{\left(n_{Z}+1\right) \log \left(\frac{1}{\epsilon_{\text {sec }}}\right)}{2 n_{X}\left(n_{X}+n_{Z}\right)}}}{1-\Delta}\right]\\
    &- f_{\mathrm{e}} n_{Z} h\left(e_{Z}\right)
    -n_{Z} \Delta-\log \frac{2}{\epsilon_{\mathrm{corr}} \epsilon_{\mathrm{sec}}^{2}},
\label{eqn:keylength}
\end{split}
\end{align}
where $\epsilon_{\mathrm{sec}}$ and $\epsilon_{\mathrm{corr}}$ are secrecy and correctness levels so that the protocol is $\epsilon$-secure if $\epsilon \geq \epsilon_{\mathrm{sec}} + \epsilon_{\mathrm{corr}}$~\cite{Yin2020}, $\Delta$ is a factor to account for the mismatch of different detector efficiencies, $n_{Z/X}$ are the number of matching and coincident $Z$ and $X$ basis detection events respectively, and $e_{Z/X}$ are the quantum bit error rates (QBERs) for each basis. The key length calculation for the $X$ basis is similar to Eq.~\ref{eqn:keylength} thus the total key length becomes $L=L_{X}+L_{Z}.$

The QBERs with a single QR node (2 QMs) as in our protocol are given by~\cite{Luong2016},
\begin{align}
\begin{split}
    e_{X}=& \lambda_{\mathrm{BSM}} \alpha_{A} \alpha_{B}\left[\epsilon_{m}\left(1-\epsilon_{\mathrm{dp}}\right)+\left(1-\epsilon_{m}\right) \epsilon_{\mathrm{dp}}\right] \\
&+\frac{1}{2}\left[1-\lambda_{\mathrm{BSM}} \alpha_{A} \alpha_{B}\right],
\label{eq:ex}
\end{split}
\end{align}
and 
\begin{align}
\begin{split}
e_{Z}=\lambda_{\mathrm{BSM}} \alpha_{A} \alpha_{B} \epsilon_{m}+\frac{1}{2}\left[1-\lambda_{\mathrm{BSM}} \alpha_{A} \alpha_{B}\right].
\label{eq:ez}
\end{split}
\end{align}
Here $\lambda_{\mathrm{BSM}}$ is a parameter that quantifies the ideality of the BSM and it is related to the BSM fidelity $F_{\mathrm{BSM}}=\sqrt{3 (\lambda_{\mathrm{BSM}+1})/{4}}$; $\alpha_{k}$ is the probability of a real detection event in ground station $k$; $\epsilon_{m}$ is the misalignment error that also includes the source infidelity and $\epsilon_{dp}$ is total dephasing during the storage in memories which depends on the individual memory errors $e_m$ (details for both single- and double-quantum memory cases are given in Appendix~\ref{appendix:rates}). Ensemble based memories that we consider in this work have been shown to preserve the phase, independent of the storage time~\cite{Staudt2007, Gundogan2013, Ma2021}. This is since re-emission of the stored information relies on rephasing of these excitations~\cite{Afzelius2009}, any dephasing will result in lower operation efficiency while maintaining high fidelity. We assume a memory efficiency of $\sim 60\%$ at 90~minutes following the observed $T_2 = 6$~hours in a Europium doped crystal~\cite{Bland-Hawthorn2021, Zhong2015} (see Sec.~\ref{sec:memory}) and $\lambda_{\mathrm{BSM}}=0.98$~\cite{Luong2016, Trenyi2020} corresponding to $F_{\mathrm{BSM}}=0.9925$. We further assume two identical passes over ground stations A and B each of $240$~s duration and without memory constraints. Consequently, we will determine the required memory capacity as a result. We also do not assume a particular orbit, apart from being consistent with the overpass times and channel losses considered and achievable with realistic transmitter and receiver apertures. 

Figure~\ref{Fig2}a shows the key rate $R$ per \emph{received} pair for transmission periods of $240$~seconds per ground station, as a function of average single channel loss ($\eta_{ch}$) and different memory dephasing values $e_{m}$. Since the key rate only depends on the average channel loss, this figure serves as a guide for scenarios with a given total channel loss~\footnote{The exact channel loss will depend on the specifics of the system, e.g. transmitter and receiver aperture, wavelength, overpass geometry, but for the purposes of a comparison between the two different memory configurations, these have been abstracted away as they should not alter the relative performance.}. We assume an EPS rate $s=5$~MHz (mainly limited by the narrow memory bandwidth), $\epsilon_{m}=2\%$, and a tight $\epsilon_{corr}=\epsilon_{sec}=5\times10^{-12}$~\footnote{According to Ref.~\cite{Tomamichel2017, Lim2021}, when corrected, this corresponds to a larger security parameter of around $10^{-9}$.} (see Appendices). We see that up to $\eta_{ch}=42$~dB can be tolerated with 2 ideal QMs ($e_{m}=0\%$). For $e_{m}=5\%$ one can generate finite keys up to $37.5$~dB single-channel loss. For perspective, Ref.~\cite{Ma2021} reported $e_{m}\sim 7.5\%$ for 1~hour storage time with classical pulses. Fig.~\ref{Fig2}b shows that finite key lengths ($L=L_X+L_Z$) of $>10^4$ can be obtained with an average single channel loss of $30$~dB, such $\eta_{ch}$ values have been report in~\cite{Yin2020}.

\begin{figure}[tbh!]
\centering
  \begin{tabular}{@{}cccc@{}}
    \includegraphics[width=0.95\columnwidth]{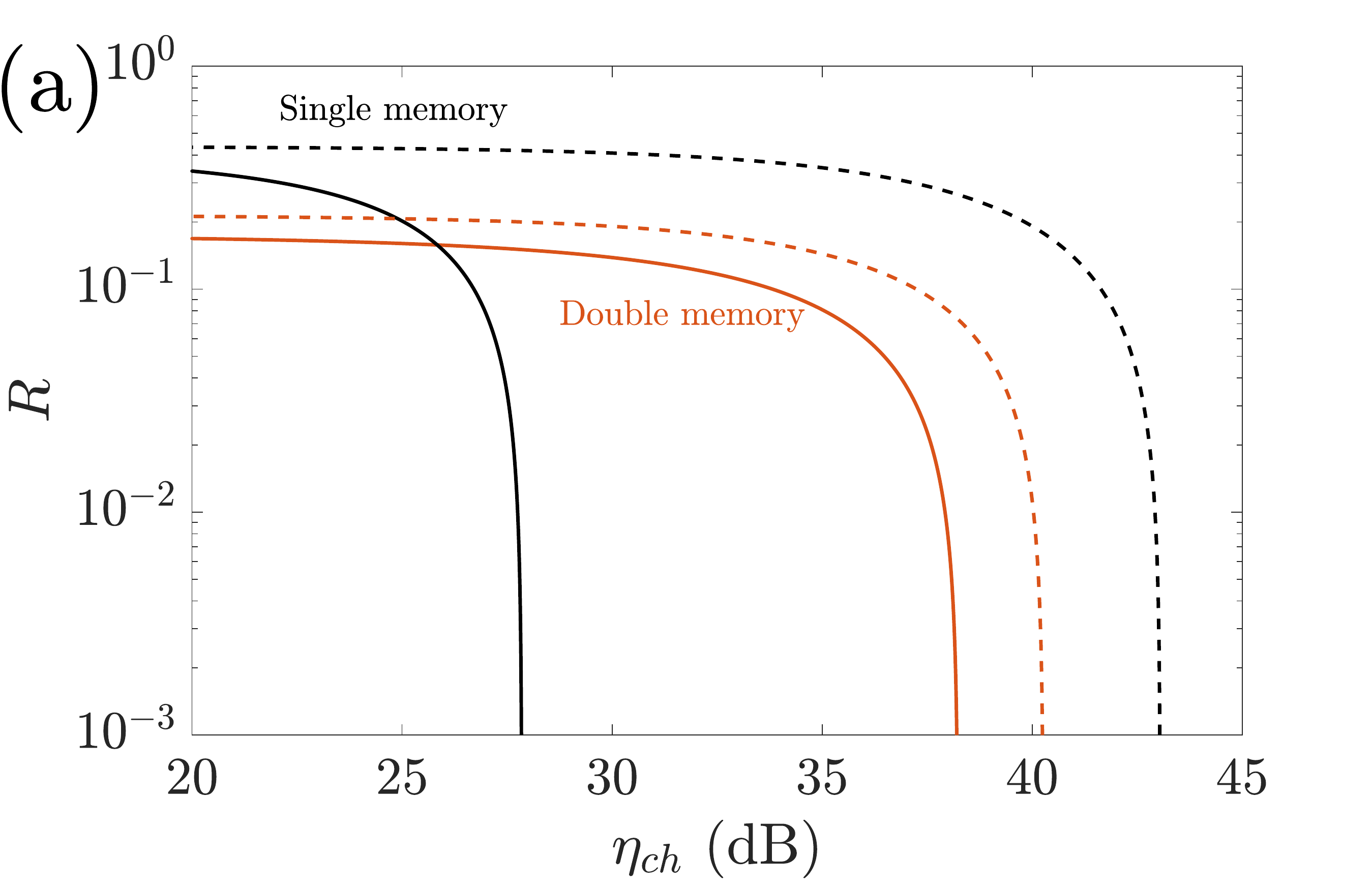}\\
    \includegraphics[width=0.95\columnwidth]{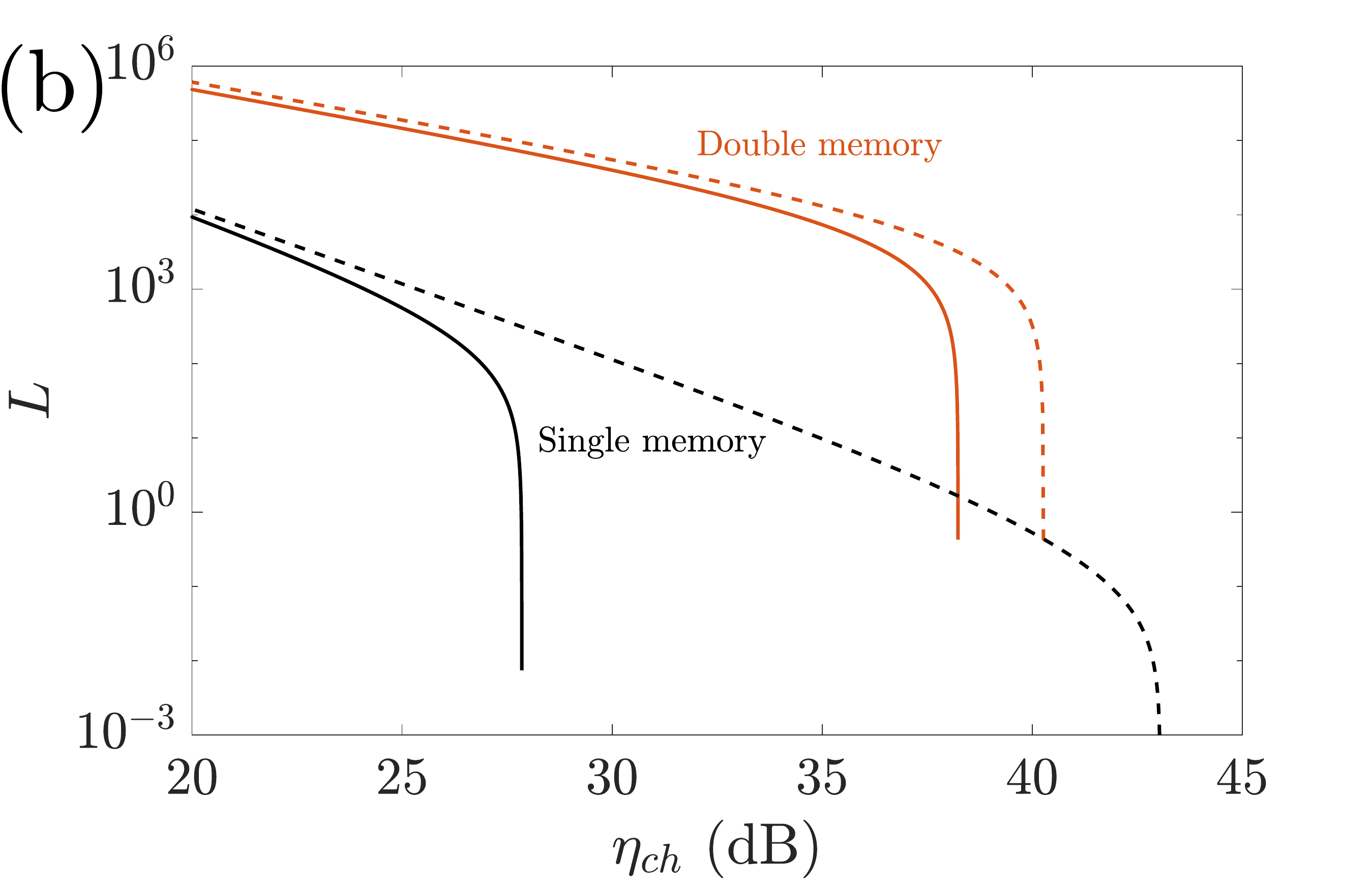}\\
    \includegraphics[width=0.95\columnwidth]{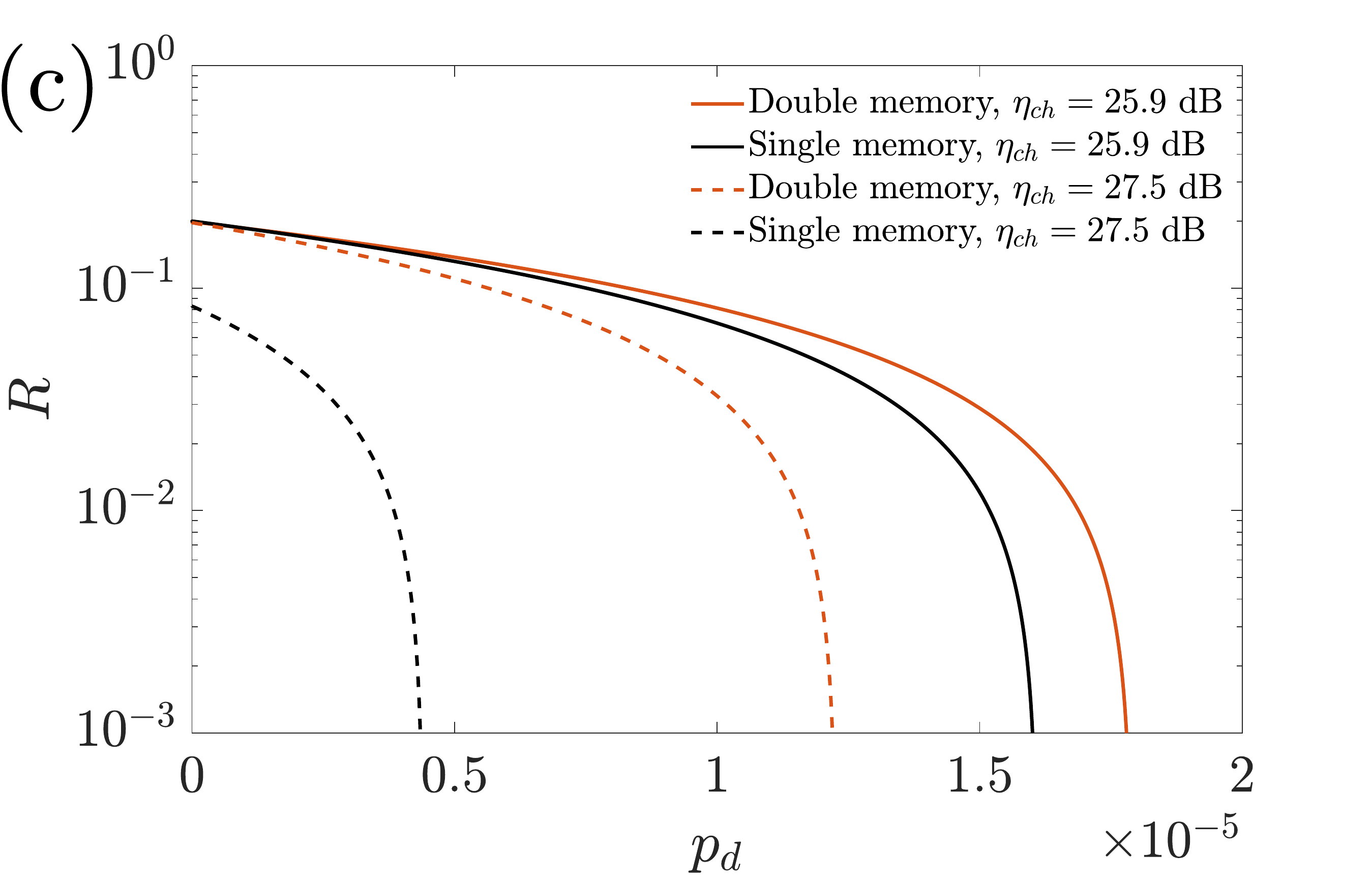}\\
  \end{tabular}
  \caption{Comparison of 1-QM and 2-QM Key Generation. a) Finite key rate per received pair (R) for $e_{m}=5\%$ (solid/dashed curves finite/asymptotic key rate); b) total finite key length (L) as a function of average single channel loss ($\eta_{ch}$) for $e_{m}=5\%$ (solid/dashed curves finite/asymptotic key length). Solid red curves in a) and b) are the same as the yellow curves in Fig.~\ref{Fig2}. c) Finite key rate per received pair versus total incoherent noise for fixed channel losses (crossover between finite key rates in a) is $\eta_{ch}=25.9$~dB).\label{Fig3}}
\end{figure}

Compared with the 1-QM scheme~\cite{Wittig2017}, the second QM provides a marked advantage in the finite key setting (Fig.~\ref{Fig3}). With a single QM, finite size effects becomes significant after $\eta_{ch}=\sim26$~dB and the maximum tolerable average loss is $28$~dB, beyond which secure key generation is not possible in a single set of overpasses. This also means that the 1-QM scheme would not produce any finite key with channel losses such as those reported in Ref.~\cite{Yin2020}. Fig.~\ref{Fig3}b explicitly demonstrates the advantage of the two-memory scheme, with higher loss tolerance and with orders of magnitude higher secure key lengths due to the repeater effect on the second downlink indicated by the lower slope of the curve for the 2-QM case.

The 2-QM protocol can achieve a finite key rate that approaches its asymptotic limit even for high channel loss and contact times of only a few minutes. In contrast, the 1-QM scheme shows a $15$~dB gap between the maximum tolerable loss and finite key limit due to the greatly reduced number of received pairs, without the repeater effect of the second QM, imposing a severe finite block size penalty. In the asymptotic limit, however, the single-memory scheme could in principle tolerate higher losses than the two-memory case due to a reduction of errors from the absence of a non-perfect BSM ($\lambda_{\mathrm{BSM}}=1$ in the 1-QM case effectively) and the additional dephasing in QM2 (Appendix~\ref{appendix:rates}). We also note that in the 2-QM case, the BSM maximum success rate of 50\% (assuming passive ancilla-less static linear optics~\cite{zaidi2013beating}) halves the number of received pairs, hence the implementation of deterministic entanglement swapping BSMs~\cite{riebe2008deterministic} could provide a tangible improvement in finite key generation and increasing loss tolerance due to longer blocks and better finite statistics.

Fig.~\ref{Fig3}c shows the effect of incoherent detector clicks, $p_d$, on the key rate for $\eta_{ch}=25.9$~dB (Appendix~\ref{appendix:pd}), the crossover point in Fig.~\ref{Fig3}a between the two schemes in the finite key setting. The two-memory scheme is more resilient to noise, despite the additional errors introduced by the BSM/entanglement swapping and second memory required. This is due to the much larger block size achievable with 2-QMs leading to lower statistical uncertainties, hence tighter bounds in Eq.~\ref{eqn:keylength}. The sensitivity on the single channel loss is also illustrated: the dashed curves for $\eta_{ch}=27.5$~dB show that the double memory scheme has more than three times better noise tolerance than its single memory counterpart.

\section{Memory architecture}\label{sec:memory}

\noindent
For both 1-QM or 2-QM protocols, the primary memory (QM1) needs to have: i) high multimode storage capacity, $N$; ii) long storage times. For the 2-QM protocol, $N=4(n_Z+n_X)$,  where the factor of 4 arises from two contributions: a) 50\% basis mismatch between ground stations A and B; b) assumed $50\%$~BSM success limit. The storage time should be comparable to the orbital period ($\sim90-100$~min for LEO). Though half an orbital period would be sufficient in principle, it may take several orbits to overfly one ground station and then the other, hence longer storage times are desirable. Mode capacities of $N\sim200$ have recently been demonstrated in several experiments: cold atoms~\cite{Pu2017} with spatial multiplexing; rare-earth ion doped (REID) crystal memories with temporal and spectral multiplexing~\cite{Seri2019}.

REIDs have the long memory lifetime and multimode capacity required by our protocol. Due to the optically active $4f$ electronic orbitals being located within totally filled $5s$ or $5d$ shells~\cite{Goldner2015}, this shields them from external disturbances and results in very sharp optical lines connected to long-lived ground states. Their large inhomogeneous broadening ($\gamma_{\text{inh}}$) can also be used to realise memory protocols based on rephasing of the stored excitations using the atomic frequency comb (AFC) protocol~\cite{Afzelius2009}. Europium REIDs have an hyperfine lifetime of around three weeks~\cite{Konz2003, Lauritzen2012} with demonstrations of a coherence time of 6~hours~\cite{Zhong2015}, and recently the coherent storage of bright pulses for $>1$~hour~\cite{Ma2021}. Additionally, recent advances in developing such QMs with built-in fibre couplers~\cite{Rakonjac2021b, Liu2022} would enable enhanced coupling between the optical free space and memory modes.

The AFC protocol is a promising candidate to satisfy the strict multimode requirements of our scheme. AFC relies on creating a comb shaped absorption profile by optical pumping within the inhomogeneous profile of the ensemble. It was shown that the number of temporal modes that can be stored with this protocol is $\sim N_{AFC}/6$, where $N_{AFC}$ is the number absorption peaks within the comb~\cite{Afzelius2009}. With a bandwidth of a few MHz, we can expect to create an AFC that could store $N_t\sim10^2$ temporal modes with $N_{AFC}\sim600$~\cite{Su2022}. REIDs are also suitable for large spectral multiplexing~\cite{Sinclair2014, Seri2019} capability due to the inhomogeneous broadening they possess: up to $N_f\sim10^3$ spectral modes can be stored within an optical transition with $\gamma_{\text{inh}}\sim10$~GHz. Finally, laser waveguide writing techniques~\cite{Corrielli2016, Su2022} may allow creating $N_s\sim100\times100$ arrays of spatial modes within a single crystal, thus putting the total number of available modes to $N_{\text{Mem}}=N_t \times N_f \times N_s\sim10^9$. This hypothetical value is well beyond the required capacity of $N_{\text{QM1}}\sim2\times10^6$ for 30~dB average downlink single channel loss in combination with $\eta_{mem}=0.6$ and $\eta_{det}=0.8$. We note that for LEO ranges and $s=5$~MHz, we require $N_{\text{QM2}}$ $\sim 10^5$ modes to act as a buffer taking into account the propagation delay and the need to wait for the click/no-click signal from the ground station.

These types of quantum memories require cryogenic operation at temperatures $<4$~K and external magnetic fields~\cite{McAuslan2012}. Although this is one of the technical challenges for deploying such devices in space, recent efforts to develop satellite-borne cryostats for quantum optical applications are promising.~\cite{Olson2014, You2018, Dang2019}. Highly efficient, low-noise single photon counters are also available. We should note that detector timing jitter ($<1$~ns) is not a significant factor as the signal photons would need to be around $\tau=200$~ns long due to narrow memory linewidths. We should note that QM2 can be implemented with the same memory protocol which would also ensure the indistinguishability of photons for the BSM. The storage time and multiplexing requirements for QM2 are greatly relaxed compared with QM1. In Appendix~\ref{sec:GEO} we also compare the annual performance of the 2-QM protocol ($s=5$~MHz) with a non-memory assisted continuously operating simultaneous dual-downlink from geostationary orbit ($s=1$~GHz).

\section{Conclusions}

We propose a new quantum communication protocol that physically transports stored qubits in an ULL QM on an orbiting satellite, together with a second (non-ULL) QM to substantially enhance entanglement distribution over long distances. This protocol dramatically reduces system complexity of global quantum networks by taking advantage of two different paradigms, i.e. quantum repeater behaviour and physically moving qubits, eliminating the need to coordinate orbiting strings of QR satellites and multiple optical links simultaneously. Using two QMs instead of one significantly increases the maximum tolerable channel loss while reducing the required multimode capacity from $\sim10^8$~\cite{Wittig2017} to $\sim10^5$ despite additional errors from the second QM, a non-ideal BSM, and 50\% BSM outcome inefficiency in the 2-QM case. Recent progress in QMs indicates that the necessary storage time and multimode capacity should be achievable in the near future.

These results could be improved by utilising more recent finite key calculations that specifically address space-based QKD scenarios. The secure key lengths may be increased by approximately $\sim 10\%$ or else smaller block sizes could be used~\cite{Lim2021}. Wavelength division multiplexing may allow increased rates at which entangled pairs can be sent through the space-Earth quantum optical channel despite QM linewidth limitations. Finally, ULL QMs in orbit may also serve as useful probes to investigate the intersection of quantum physics and general relativity~\cite{Barzel2022} and enable human-assisted Bell tests across Earth-Moon distances~\cite{Gundogan2021b, Lu2022, Mol2023}.

There are a number of extensions to this work, including: evaluating the trade-off between the data block sizes and the average overpass loss to maximise the finite key rate could provide requirements and specifications on quantum memory capacity and performance to support this single-node quantum repeater scheme. It may also help determine current hardware limitations to extending the protocol performance. Further, exploring the impact of different satellite overpasses on the secure key generation would elucidate the robustness of our single satellite scheme against loss and overpass duration, and inform constellation design.

\section{Acknowledgements}

We thank S. Wittig for bringing the earlier work~\cite{Wittig2017} to our attention during discussion at an early stage of this work and E. \.{I}mre for useful discussions on satellite communication techniques.  MG and MK acknowledge the support from DLR through funds provided by BMWi (OPTIMO, No.~50WM1958 and OPTIMO-II, No.~50WM2055), MG is further supported by funding from the European Union's Horizon 2020 research and innovation programme under the Marie Sk{\l}odowska-Curie grant agreement No.~894590 (QSPACE).  DKLO is supported by the EPSRC Researcher in Residence programme (EP/T517288/1). JSS, and DKLO acknowledge the travel support by the EU COST action QTSpace (CA15220), the UK Space Agency (NSTP3-FT-063 ``Quantum Research CubeSat'', NSTP Fast Track ``System Integration \& Testing of a CubeSat WCP QKD Payload to TRL5''), EPSRC Quantum Technology Hub in Quantum Communication Partnership Resource (EP/M013472/1) and Phase 2 (EP/T001011/1), and Innovate UK (EP/S000364/1). This work was supported by the EPSRC International Network in Space Quantum Technologies INSQT (EP/W027011/1).

\appendix

\section{Key rate calculation details}
\label{appendix:rates}

Satellite based quantum communications present several challenges with respect to its fibre-based counterpart. These are mainly: high, dynamic channel loss and the relatively short contact time ($\sim$min for LEO) between the sender and receiver stations. With these constraints in mind, we also take into account the finite key effects on the final, generated secure key lengths. We should stress that we base our calculations on the average single channel loss, $\eta_{ch}$, which can be achieved with many different combinations of orbital (altitude, latitude etc.) and optical parameters (beam divergence, sender/receiver telescope apertures), for instance. 

The final secure key length, $L_{X,Z}$, is given in Eq.~\ref{eqn:keylength}. We see from Equations~\ref{eq:ex} and \ref{eq:ez} that the physical factors that contribute to the QBERs, $e_X$ and $e_Z$ are $\lambda_{BSM}$, $\alpha_{k}$, $\epsilon_{dp}$ and $\epsilon_{m}$. Among these only the memory dephasing, $\epsilon_{dp}$ contributes to error in the $X$ basis. Here, $\alpha_{k}$ is the probability of registering a real click in ground station $k$~\cite{Luong2016}. This probability then decreases with any incoherent (noise) click at the detector and depends on total detection probability $\eta$:
\begin{equation}
    \alpha(\eta) = \frac{\eta(1-p_d)}{1-(1-\eta)(1-p_d)^2},
\end{equation}
where $\eta=\eta_{ch}\eta_{det}\eta_{mem}$ where $\eta_{ch}$ is average channel transmission; $\eta_{det}$ is detection efficiency and $\eta_{mem}$ is the combined memory write-in and read-out efficiency. The probability of any incoherent click on the detector during the detection time window is $p_d = \eta_{ch}p_{n}+p_{bg}+p_{dc}$ where: $p_{n}$ is the noise added by memory when there is no input pulse; $p_{bg}$ is the background noise probability; and $p_{dc}$ is detector dark count probability. These contributions are further detailed in the next section. 

The efficiency mismatch between different detectors is accounted by $\Delta$ in Eq.~\ref{eqn:keylength}. Adding a filter with small attenuation, $\delta_{i}$, to the $i$th detector would result in $\Delta = 1-1/(1+\delta)$ ~\cite{Yin2020}. In this work we assume $\Delta\sim2\%$.

The memory dephasing term, $\epsilon_{dp}$, combines the dephasing experienced by both memories and is expressed as~\cite{Luong2016}
\begin{equation}
    \epsilon_{dp} = e_{m 1}\left(1-e_{m 2}\right)+\left(1-e_{m 1}\right) e_{m 2}
\end{equation}
with $e_{m 1,2}$ depicting the dephasing experienced in individual memories which is assumed to be independent of the storage time as explained in the main text. With this, $\epsilon_{dp}$ becomes $2 e_{m}\left(1-e_{m}\right)$ for two- and $e_{m}$ for single-memory case, where $e_{m}$ is the memory dephasing. With these, Eq.~\ref{eq:ex} becomes,
\begin{align}
\begin{split}
    e_{X}=&\lambda_{\mathrm{BSM}} \alpha_{A} \alpha_{B}[\epsilon_{m}\left(1-2 e_{m}\left(1-e_{m}\right)\right)+\\
  & 2 e_{m}\left(1-e_{m}\right) \left(1-\epsilon_{m}\right)] +\frac{1}{2}\left[1-\lambda_{\mathrm{BSM}} \alpha_{A} \alpha_{B}\right]
\label{eq:ex_app}
\end{split}\\
e_{Z}=&\lambda_{\mathrm{BSM}} \alpha_{A} \alpha_{B} \epsilon_{m}+\frac{1}{2}\left[1-\lambda_{\mathrm{BSM}} \alpha_{A} \alpha_{B}\right]
\label{eq:ez_app}
\end{align}
for the two-memory and,
\begin{align}
    e_{X}=&\alpha_{A} \alpha_{B}[\epsilon_{m} \left(1-e_{m}\right) + e_{m}\left(1-\epsilon_{m}\right)]
\label{eq:ex_appB}\\
e_{Z}=&\alpha_{A} \alpha_{B} \epsilon_{m}+\frac{1}{2}\left[1-\alpha_{A} \alpha_{B}\right]
\label{eq:ezB}
\end{align}
for the single-memory scenarios, respectively.

\newcommand*{\tabindent}{ \hspace{-1mm}}
\newlength{\thickarrayrulewidth}
\setlength{\thickarrayrulewidth}{2.1\arrayrulewidth}
\setlength{\tabcolsep}{4pt}
\begin{table}[t!]
  \centering
  \caption{\label{tab:parameters}Reference simulation parameters. A transmission duration of 240~s over an OGS corresponds to typical LEO overpasses. Memory noise, $p_n$ corresponds to the probability a noise photon is emitted per storage trial. A value of $p_n = 1 \times 10^{-3}$ is consistent with current REID memories~\cite{Gundogan2015, Ortu2022}. The assumed background count probability is consistent with nighttime operation. Daylight operations may be possible with extremely aggressive spectral filtering, diffraction-limited adaptive-optics enabled single-mode coupling, and favourable clear sky conditions and overpass geometry, though such low values may not be achievable in the majority of circumstances. However, as can been seen in Fig.\ref{Fig3}c, the 2-QM protocol can operate with higher $p_d$ values for achievable single channel losses $\eta_{ch}$.}
  \begin{ruledtabular}
  \begin{tabular}{m{4.8cm}m{1.4cm}m{1.6cm}}
    \textbf{Parameter} 				& \textbf{Symbol}			& \textbf{Value} 	\\
    \hline
    \footnotesize EPS rate 			        	&	$s$						& 5 MHz			\\
    \footnotesize Transmission period				&	$T$						& 240 s			\\
    \footnotesize Average channel loss			&	$\eta_{ch}$				& \textit{See text}	\\ 
    \footnotesize Secrecy parameter		        &	$\epsilon_{sec}$		& $5\times10^{-12}$ 	\\   
    \footnotesize Correctness parameter			&	$\epsilon_{corr}$		& $5\times10^{-12}$ 	\\  
    \footnotesize Memory noise probability					&	$p_{n}$					& $1\times10^{-3}$ \\  
    \footnotesize Background count probability			&	$p_{bg}$				& $6.4\times10^{-7}$ 	\\  
    \footnotesize Detector dark count probability			&	$p_{dc}$				& $1\times10^{-7}$ 		\\
    \footnotesize Temporal Window                 &   $\tau$                  & $200$~ns  \\
    \footnotesize Memory efficiency				&	$\eta_{mem}$			& $0.6$		\\   
    \footnotesize Detection efficiency			&	$\eta_{det}$			& $0.8$ 		\\     
    \footnotesize Memory dephasing				&	$e_{m}$                 & $0-0.05$		\\
    \footnotesize Detector imbalance              &   $\Delta$                & $0.02$  \\
    \footnotesize BSM ideality            & $\lambda_\mathrm{BSM}$  & 0.98 \\
  \end{tabular}
  \end{ruledtabular}
\end{table}%
%


\section{Incoherent clicks}
\label{appendix:pd}

Incoherent clicks degrade the maximum tolerable loss. These arise from: a) Detector dark counts, $p_{dc}$. Dark count rates of around 10~Hz is readily achievable with detectors at room temperature whereas $<1$~Hz is possible with cryogenically cooled superconducting single photon detectors. b) Noise coming from memory operation, $p_{n}$. Long-lived QMs usually add noise to the output signal due to a combination of several factors such as imperfect preparation of the memory, leakage from the strong control pulses and resonant four wave mixing noise. The probability of emitting a noise photon per storage trial, $p_n$, is around $10^{-3}$ for REID QMs~\cite{Gundogan2015, Ortu2022} whereas $\sim10^{-4}$ has been achieved with laser cooled gases~\cite{Heller2022}. c) Background light, $p_{bg}$. Detailed treatment of background light under different conditions (total dark skies, full moon, daylight etc.) is given elsewhere~\cite{Er_long2005, Wallnofer2022}.
The use of adaptive optics can compensate for turbulence to allow for a diffraction limited field of view and single mode coupling, hence greatly improving background light rejection, permitting daylight satellite QKD even at $800$~nm~\cite{gruneisen2021adaptive}. Background light can be reduced further with even narrower spectral filters (in principle, approaching the bandwidth of stored photons or lower~\cite{Beavan2013, Gundogan2015}), though the Doppler shift due to satellite's velocity along the line-of-sight would need to be compensated. Furthermore, one can design the photon pair source such that photons sent to the ground stations would lie within a Fraunhofer line to enable daylight operation~\cite{Abasifard2023}. This option would not be directly possible in a 1-QM scheme as the stored photons would later have to be transmitted to ground station B, unless wavelength conversion was employed.

\section{Comparison with a QKD from geostationary orbit (GEO)}\label{sec:GEO}

\begin{figure}[h]
    \includegraphics[width=\columnwidth]{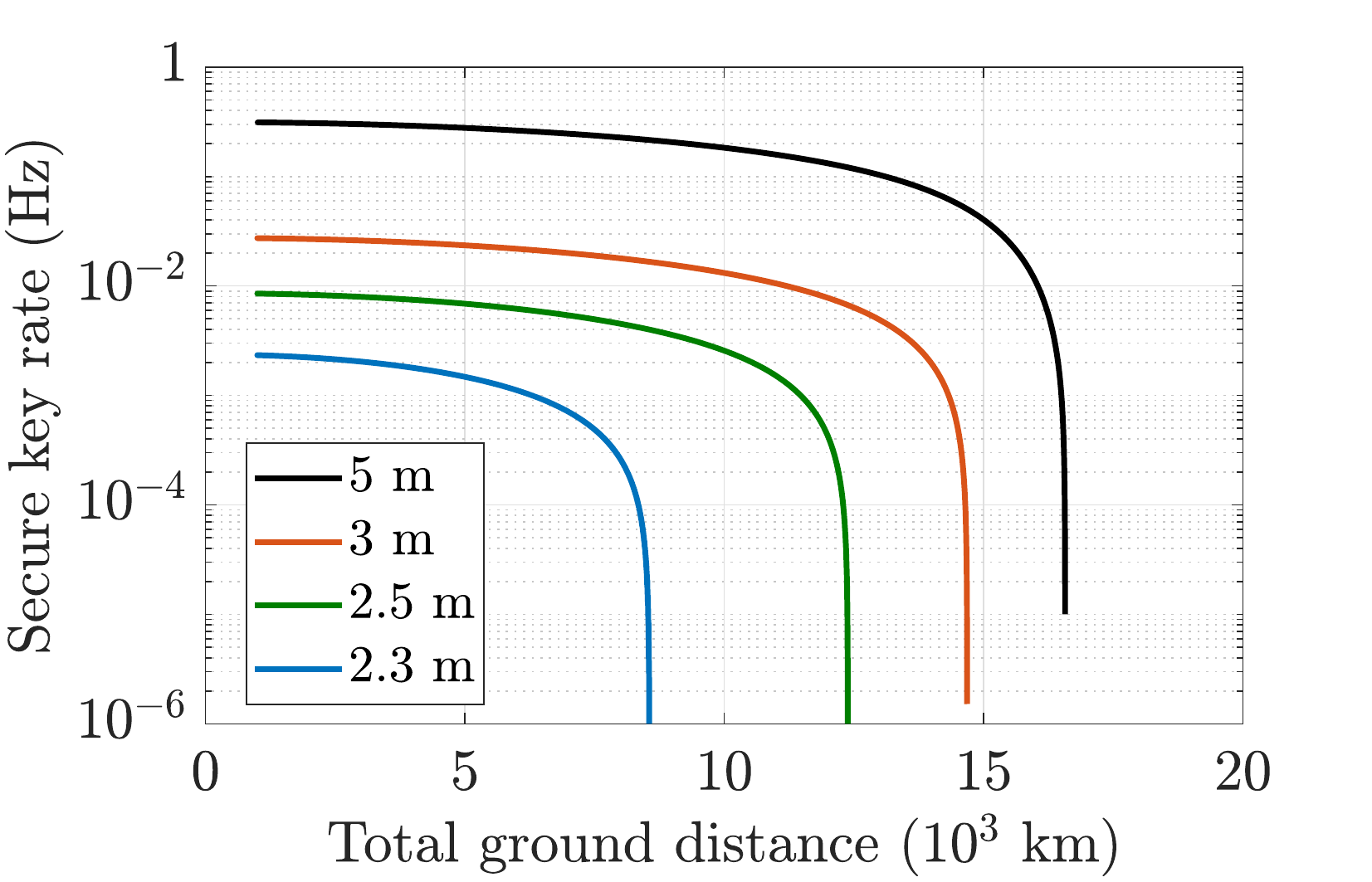}
    \caption{Achievable key rates from a GEO with an ent-QKD protocol without any QM with different receiver telescope diameters. Other parameters are given in Appendix.~\ref{sec:GEO}.}
    \label{fig:GEO}
\end{figure}

Entanglement could be distributed globally with the use of high altitude satellites simultaneously transmitting photon pairs. This requires the satellite to be sufficiently high to be in line-of sight of the two ground stations at the same time.
Following Refs.~\cite{Ma2007, Gundogan2021}, in this section we present calculations of entanglement-based QKD (ent-QKD) without any QMs from a satellite in GEO (see Ref.~\cite{dirks2021geoqkd} for a similar analysis) at an altitude of $36\times10^3$~km with different receiver telescope diameters. We assume a source rate of 1~GHz, sender telescope diameter of 0.3~m, $p_{d}=10^{-6}$, wavelength of 852~nm (atmospheric transmissivity at 852 nm is around $25\%$ higher than it is for 580~nm~\cite{Gundogan2021}) and beam divergence of 5~$\mu$rad. We follow our earlier work~\cite{Gundogan2021, Wallnofer2022} for the channel modelling. This calculation does not include pointing and tracking errors or turbulence effects thus represents a best case link loss scenario.

With this, we see that losses prevent achieving fully global distances, even with 2.5~m diameter receiver telescopes. In order to compare this GEO scenario with the presented protocol in this manuscript one needs to calculate the number of flyovers over a given pair of ground stations. Along these lines, in Ref.~\cite{Wittig2017} a hypothetical link between Adelaide and Brussels (total distance $15.9\times10^3$~km) is considered for which the authors find 1257 flyover pairs within a year. According to Fig.~\ref{Fig2}b, each of these passes can generate finite key lengths up to $10^4$ with average single channel loss of $35$~dB. Our analysis shows that deploying a single satellite equipped with a ULL QM using the protocol in this manuscript would result in a performance gain of $\sim2.5\times10^2$ over the GEO scenario per Fig.~~\ref{fig:GEO}.


%

\end{document}